# NeoSySPArtaN: A Neuro-Symbolic Spin Prediction Architecture for higher-order multipole waveforms from eccentric Binary Black Hole mergers using Numerical Relativity

**A Vibho[1, 2], A. A. Bataineh[2]**

[1] College of Science and Mathematics, Norwich University, 158 Harmon Drive, Northfield, VT 05663, United States

[2] Center for Artificial Intelligence, Department of Electrical and Computer Engineering, David Crawford School of Engineering, Norwich University, 158 Harmon Drive, Northfield, VT 05663, United States.

avibho@stu.norwich.edu (corresponding author), aalbatai@norwich.edu

## Abstract

The prediction of spin magnitudes in binary black hole and neutron star mergers is crucial for understanding the astrophysical processes and gravitational wave (GW) signals emitted during these cataclysmic events. In this paper, we present a novel Neuro-Symbolic Architecture (NSA) that combines the power of neural networks and symbolic regression to accurately predict spin magnitudes of black hole and neutron star mergers. Our approach utilizes GW waveform data obtained from numerical relativity simulations in the SXS Waveform catalog. By combining these two approaches, we leverage the strengths of both paradigms, enabling a comprehensive and accurate prediction of spin magnitudes. Our experiments demonstrate that the proposed architecture achieves an impressive root-mean-squared-error (RMSE) of 0.05 and mean-squared-error (MSE) of 0.03 for the NSA model and an RMSE of 0.12 for the symbolic regression model alone. We train this model to handle higher-order multipole waveforms, with a specific focus on eccentric candidates, which are known to exhibit unique characteristics. Our results provide a robust and interpretable framework for predicting spin magnitudes in mergers. This has implications for understanding the astrophysical properties of black holes and deciphering the physics underlying the GW signals.

**Keywords:** black hole physics, methods: numerical, gravitational waves, transients: black hole - neutron star mergers, software: data analysis,

## I. Introduction

In his revolutionizing paper, Princeton physicist John Wheeler coined the term *black hole* for a class of stellar remnants formed by the collapse of massive stars under their own gravity (Ruffini and Wheeler, 1971). Though the concept had been conceived by John Michell in 1783 and revisited and rediscovered by several pathbreaking physicists – Oppenheimer, Cavendish, and Einstein to name a few – Wheeler's coining of the term began what has now become a century of categorization, subdivision, and search for variants such as supermassive, stellar-mass, and the most prominent in candidacy for dark matter, through which the Standard Model might be revised – primordial black holes (Einstein, 1914; Hawking, 1976).

Through a plethora of theoretical and observational experiments, the scientific community today knows not only of the existence but of the very strange behavior of black hole and neutron star systems, and binary systems now known to form the center of galaxies such as our own with Sagittarius A* and S2 (Borka et al., 2021).

Further, with the detection of the very first gravitational waves through the Laser Interferometer and Gravitational Wave Observatory (LIGO) followed by the release of the first *image* of a black hole through the Event Horizon Telescope (Akiyama et al., 2019), Einstein's theory of General Relativity has been proven as ground truth with increasing confidence (Abbott et al., 2016). These milestones then raised a question in visualizing black hole, neutron star, and black hole-neutron star mergers in the most extreme circumstances, paving the way for post-Newtonian calculations of General Relativity (GR) to be extended to a new era of computational astrophysics namely, Numerical Relativity (NR) ( Lehner, 2001; Baumgarte and Shapiro, 2010; Shibata, 2015).

Since the advent of NR, numerous software packages have helped simulate these extreme circumstances through a combination of General Relativity formalisms, magnetohydrodynamics, and recently artificial intelligence architectures, capturing relativistic frameworks and mathematical relationships that might not be apparent otherwise (Huerta *et al.*, 2018; Whittaker *et al.*, 2022; Khan, Huerta and Zheng, 2022). The two formalisms: Arnowitt-Deser-Misner (ADM) that interprets GR through the Hamiltonian school of thought (Corichi and Núñez, 2022) and the Baumgarte-Shapiro-Shibata-Nakamura (BSSN) that introduces auxiliary variables in the Hamiltonian, ADM formalism (Mertens, Giblin and Starkman, 2016) have been integrated within simulation code packages. Einstein Toolkit, Spectral Einstein Code, and ENIGMA are mere examples of powerful and enabling NR packages that have resulted in a plethora of simulations and post-simulation analyses in the past decade (Tsatsin *et al.*, 2011; ZILHÃO and LÖFFLER, 2013; Haas *et al.*, 2016).

The realm of higher-order multipole waveforms of quasi-circular and/or eccentric binary black hole mergers has been investigated in great depth using artificial intelligence and NR simulations though the Einstein Toolkit (Rebei *et al.*, 2019). We embark on an exploration of this realm with a Neuro-Symbolic Approach, aiming to develop an artificial intelligence framework where we reinforce the optimum flow-of-control obtained through Symbolic Regression and integrate it with a Neural Network model with the highest accuracy trained on waveform characteristics to predict the spin of these black-hole and neutron star mergers.

In doing so, we attempt to take the first step in obtaining an advanced architecture that we could further implement and obtain equations of state relationships – polytropic and parametric – for NR simulated extreme merger systems as a way of narrowing our search space for potential dark matter candidates among the class of primordial black holes.

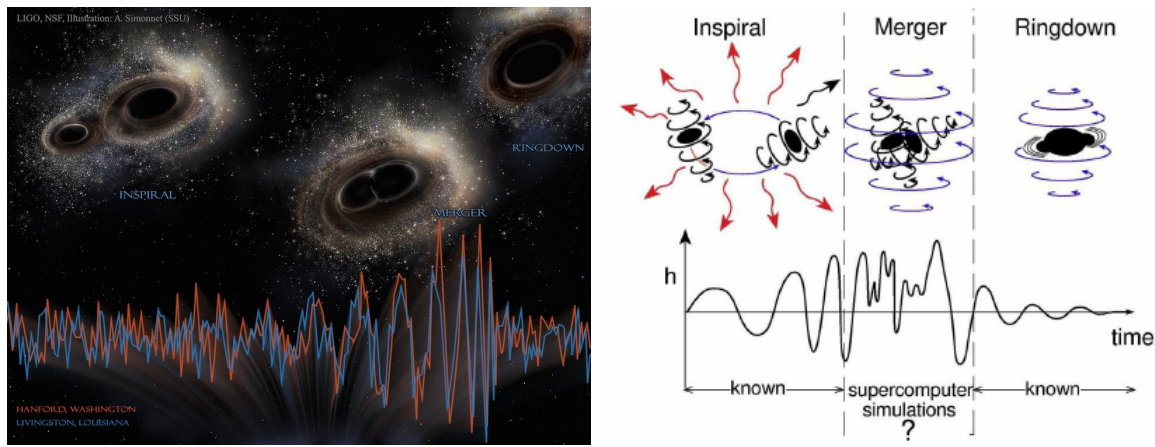

***Figure: 1 - Illustrations of the inspiral-merger-ringdown phase of hypothesized primordial black hole binary systems** (Schutz, 2006; Grizzle, 2016).*

We therefore, organize the paper beginning with Data and Methodology in Section II, Results in Section III, a discussion of our research process, scope, setbacks, and interpretations in Section IV, concluding with what our work could mean in Section V with scope for future work, followed by Author Contributions, Acknowledgements, and References in Sections VI through VIII respectively.

## II.    Data and Methodology:

*Dataset*

We use the Simulating eXtreme Spacetimes (SXS) waveform catalog of binary black hole, neutron star, and black hole-neutron star mergers produced using Spectral Einstein Code (SpEC) (Boyle *et al.*, 2019). The catalog of multi-modal gravitational wave modes with the dominant l = m = 2 mode provides simulated waveforms with mass ratios between 1 and 10 alongside 72 waveform characteristics, including those at reference time, merger, remnants, and initial configurations before the simulated inspiral-merger-ringdown phase. It was especially useful for our work that the dataset had aligned, precessing, and non-spinning waveforms from mergers, giving us a broader scope of training for our spin-prediction architecture.

*Feature selection*

Our motivation behind selecting relevant features for our spin-prediction model comes from the most recent implementation of deep learning algorithms on numerical relativity simulations of higher-order multipole waveforms of eccentric, spinning black hole mergers (Rebei *et al.*, 2019). Their feature selection consisted of the four most spin-relevant characteristics: mass-ratio, mean anomaly, orbital frequency, and eccentricity. Given that our attempt is to lay the ground for a model that can obtain polytropic or parametric equation(s) of state for these waveforms, potentially finding a range and subcategory of primordial black hole dark matter candidates, we implemented a feature importance metric using scikit-learn, giving us a set of the following features for our machine learning models:

`'reference_mass_ratio'` : mass-ratio of waveforms at reference time; we restrict 1 <= q <= 5

`'reference_eccentricity'` : orbital eccentricity of the system at reference time

`'initial_ADM_energy'` : measure of total energy of the system, including at rotation, within the construct of GR.

`'reference_mean_anomaly'` : time elapsed for the system in orbit since reference rime.

`'initial_orbital_frequency'` : orbital frequency of the system before the inspiral-merger-ringdown phase

`'initial_adot'` : rate of shrinking for the system before the inspiral-merger-ringdown phase.

NOTE: reference time refers to the time at which the inspiral-merger-ringdown phase begins. The feature-importance as computed through our scikit-learn metric is presented in Figure 2.

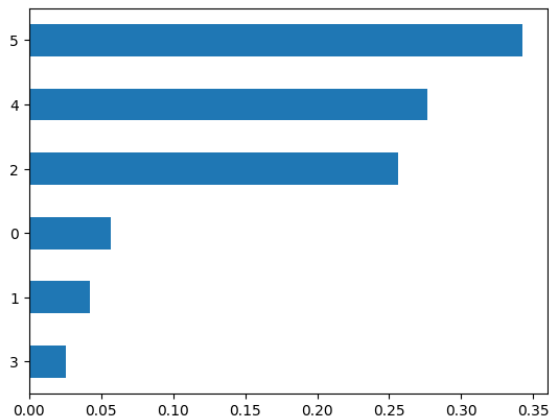

***Figure: 2 - A plot of feature-importance computed through Random Forest.***

***Machine learning models***

*Random Forest (RF)*
Given the nature of our data and the complex non-linear relationship we are trying to uncover, we choose to proceed with one of the most efficient, tree-based decision making models, Random Forest. Aiming at spin prediction, we use RandomForestRegressor. This algorithm belongs to the ensemble learning family and combines multiple decision trees to make predictions. Moreover, the Random Forest Regressor is robust against overfitting and can handle missing data and outliers, which are common challenges in astrophysical datasets (Rodriguez-Galiano *et al.*, 2015; Segal, 2017). Its ability to provide feature importance rankings also enables us to gain insights into the most influential factors contributing to the predictions in our dataset.

*Support Vector Machine (SVM)*
Support Vector Regression (SVR) is another powerful machine learning algorithm that we have chosen to analyze and model a complex black hole dataset. SVR is particularly well-suited for datasets with complex relationships and non-linear patterns. It leverages the concept of support vectors to find an optimal hyperplane that maximizes the margin between data points and the regression line. SVR is especially useful when dealing with datasets that exhibit non-linear behavior, as it can flexibly capture and model these intricate relationships. Furthermore, SVR allows for the use of different kernel functions, such as radial basis function (RBF) or polynomial kernels, which enhances its flexibility in capturing complex relationships. These kernels enable SVR to transform the data into a higher-dimensional space, where non-linear relationships can be more effectively modeled (Hastie, Tibshirani and Friedman, 2008).

*Gradient Boosting (XGB)*
We also evaluated the performance of Gradient Boosting Regressor on our dataset. While all four of our models are capable of handling non-linear relationships well, we use XGB because XGBoost also incorporates regularization techniques to prevent overfitting. It balances the trade-off between model complexity and generalization performance, ensuring robustness against noisy or irrelevant features in the dataset. This makes XGBoost particularly suitable for black hole datasets, which often contain various sources of noise and uncertainties (Friedman, 2002).

*Neural Networks and Symbolic regression: A Neuro-Symbolic Architecture (NSA)*
During the process of feature engineering, we encountered a class imbalance between aligned, precessing, and non-spinning waveforms. That being said, we tried to classify them based on the parameters we have for our regression problem, including the effective spin. We noticed a very high false positive rate for classification of nearly all unclassified waveforms as non-spinning, even with a high and near-perfect model accuracy. This was our primary source of motivation to choose Neural Networks and Symbolic Regression to pinpoint what the decision -making process was (Livni, Shalev-Shwartz and Shamir, 2014; George and Huerta, 2018; Bister *et al.*, 2021). These are both high-performing models and can 'see' patterns beyond the standard machine learning algorithms and with *rule extraction* for symbolic regression using Decision Trees, we could extract the process to later merge it with our pre-trained NN model to make a Neuro-Symbolic model that resulted in an algorithm more efficient that the sum of its parts (Hitzler and Sarker, 2022; Zhang *et al.*, 2022). We discuss our classification model in greater detail in Section IV.

When we implemented feature importance for the NSA model, we implemented *rule extraction* using Decision Trees, as mentioned above. This allowed the exact node-by-node and branch-by-branch layout of the decision making process to be extracted as a set of rules. To interpret these *rules*, we used a recursion function, feeding them back to a dataset of known values and an assigned threshold, regressing over which we could extract an optimum

'depth' for our features, as in the optimal number of and best features for our model, giving us `'initial_orbital_frequency', 'initial_ADM_energy' , and 'initial_adot'` as most efficient features for our decision making process.

We then integrated the extracted rules into a function with our MLPRegressor model to form a Neuro-Symbolic Architecture, results from which are discussed in the following section.

## III. Results:

Our models performed fairly accurately on the dataset, with root-mean-squared-errors of .14 for Random Forest and Gradient Boosting, and 0.33 for Support Vector Machines. Our Neural Network model performed at an RMSE of 033, with our Symbolic Regression model at an RMSE of 0.16. Combining the NN and Symbolic model, we achieved an RMSE of 0.06 for our Neuro-Symbolic Architecture (NSA), and an MSE of 0.003.

We present a comparative analysis of the RMSE scores of RandomForest, GradientBoosting, and Support Vector Machines in Figure 3. Our NSA model's evaluation metrics are represented in Figure 4, followed by a flowchart representation of our Symbolic Regression model in Figure 5.

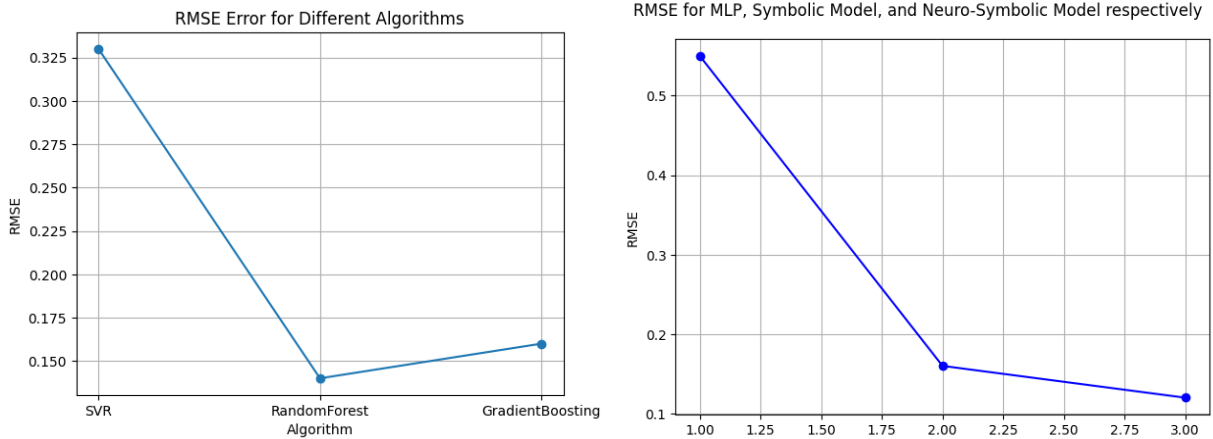

*Figures: 3 and 4 - RMSE errors for SVR, RF, XGB; RMSE for MLP, SymbolicR, and NSA.*

```
if   (initial_adot   >   -0.0)   and   (initial_adot   <=   0.0)   and
(initial_ADM_energy   <=   0.993) then response: 0.048 | based on 436
samples
if (initial_adot <= -0.0) and (initial_orbital_frequency <= 0.015)
and (initial_ADM_energy <= 0.995) then response: -0.028 | based on 327
samples
if   (initial_adot   >   -0.0)   and   (initial_adot   >   0.0)   and
(initial_ADM_energy   <=   0.995) then response: 0.329 | based on 239
samples
if (initial_adot <= -0.0) and (initial_orbital_frequency <= 0.015) and
(initial_adot <= -0.0) then response: -0.4 | based on 141 samples
if (initial_adot <= -0.0) and (initial_orbital_frequency <= 0.015) and
(initial_adot > -0.0) then response: -0.195 | based on 126 samples
if   (initial_adot   >   -0.0)   and   (initial_adot   <=   0.0)   and
(initial_ADM_energy   >   0.995)   then   response:   -0.068   |   based   on   40
samples
if   (initial_adot   >   -0.0)   and   (initial_adot   <=   0.0)   and
(initial_ADM_energy   <=   0.993)   then response: 0.366 | based on 27
samples
if (initial_adot <= -0.0) and (initial_orbital_frequency > 0.015) and
(initial_ADM_energy   >   0.995)   then   response:   -0.477   |   based   on   22
samples
```

*Figure: 5 - Extracted rules for max_depth = 3 (most important features) for the Symbolic Regression model.*

## IV.    Discussion:

This study started off as an attempt at creating a machine-learned architecture using which we could obtain equations of state parameters for the binary black hole merger waveforms in the SXS Waveform catalog. However, upon closer investigation of the missing values in the dataset, we found that out of the 2028 waveforms present, only 9 had non-null values for EoS parameters. Had we tried to predict the EoS using these existing values, the class-imbalance would have been so skewed that it would invalidate the evaluation metric against testing for false-positives, even if we implemented synthetic oversampling (Chawla *et al.*, 2002). This then led us to question the classification of spin characteristics, among which we found a sub-dataset of unclassified candidates under 'aligned', 'precessing', and 'non-spinning'. Though there existed a class imbalance here as well, we implemented Synthetic Minority Oversampling for both multiclass and binary classification, even after which we found all the unclassified candidates to be classified as non-spinning. The fact that we had legible spin values for those candidates classified as non-spinning with models having near-perfect accuracy made us question the false-positive rates, shown in Figure 6.

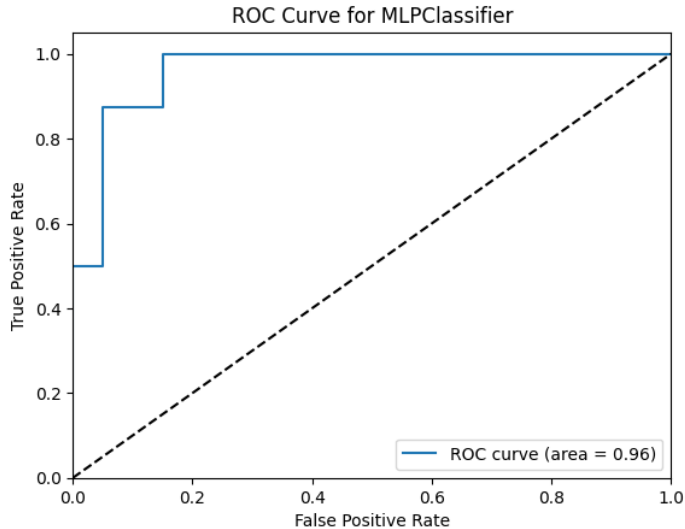

*Figure: 6 - False-positive ROC curve for our spin-classification attempt using Neural Networks.*

These two attempts brought us to the final attempt that resulted in this paper, of designing an architecture that mimics reinforcement learning (Zhang *et al.*, 2022) along with the advantages of deep learning that could correct any mispredictions and classifications for one of the most important parameters for EoS – effective spin at the inspiral-merger-ringdown phase.

Our study that follows is an attempt to focus on the EoS and this work was an important precursor to it.

## V.    Conclusion:

Through a series of attempts at correlating the spin of higher-order multipole waveforms of quasi-circular, spinning black hole mergers, we advanced our study towards creating a Neuro-Symbolic Architecture for spin prediction as a first step for obtaining equations of state parameters for these waveforms. We are interested in exploring the candidacy of primordial black holes as contenders for dark matter in lieu of the expansion of the Lambda CDM model with more information we receive through the James Webb Telescope and LIGO runs, now supplemented by Numerical Relativity simulations. With our model and errors of 0.06 and 0.003, we proceed with cautious confidence towards this approach with this paper being our first step towards a bigger leap we aim for in subsequent works.

## VI.    Author Contribution Statement:

**Amrutaa Vibho**: Conceptualization, Methodology, Data curation, Software, Writing - Original draft, editing, and final draft, Formal Analysis

**Dr. Ali Al Bataineh**: Supervision, Project Administration, Writing - Review

## VII.    Acknowledgments:

This paper would be impossible without the supervision and constant encouragement of my mentor, Dr. Ali Al Bataineh, who stood by my side through challenges with firm faith in my abilities. I am grateful for his unbridled support of my passion for computational astrophysics and artificial intelligence.

I would like to thank my parents, Bhanumathy Vishwanath and Lalithakutir Ramashesh Vishwanath for being just as excited about my interests as I have been fascinated by high-energy astrophysics, and my twin, Akshayaa Vibho, whose curiosity has been my guiding light throughout.

I acknowledge Amarnath, Drukshan, and Reine, who have been friendly faces away from home for reminding me of what I am capable of achieving, in times that I faced overwhelming doubt.

My sincere gratitude to 2LT Jacie Harlow, USMC,  my Cadet First Sergeant Shibe Couchman, and my Marine Officer Candidate mentor Destiny Sanchez for supporting me through trials and triumphs alike. I would not be here without them.